\documentclass[twocolumn]{aastex631}

\usepackage{graphicx}	
\usepackage{amsmath}	
\usepackage{booktabs} 

\newcommand{\s}{\,{\rm s}}      
\newcommand{\ps}{\,{\rm s}^{-1}}
 
\newcommand{\pyr}{\,{\rm yr}^{-1}}

\newcommand{\km}{\,{\rm km}}

\newcommand{\kpc}{\,{\rm kpc}}

\newcommand{\K}{\,{\rm K}}

\newcommand{\mas}{\,{\rm mas}}

\begin{document}

\title{Alien Type Ia supernovae from the Milky Way merger history\\ and one possible candidate: Kepler's supernova}

\correspondingauthor{Wenlang He}
\email{hewlang@smail.nju.edu.cn}
\correspondingauthor{Ping Zhou}
\email{pingzhou@nju.edu.cn}
\correspondingauthor{Eda Gjergo}
\email{eda.gjergo@gmail.com}
\correspondingauthor{Xiaoting Fu}
\email{xiaoting.fu@pmo.ac.cn}

\author[0009-0002-4427-6976]{Wenlang He}
\affiliation{School of Astronomy and Space Science, Nanjing University, 163 Xianlin Avenue, Nanjing 210023, China}
\affiliation{Key Laboratory of Modern Astronomy and Astrophysics, Nanjing University, Ministry of Education, Nanjing 210023, China}

\author[0000-0002-5683-822X]{Ping Zhou}
\affiliation{School of Astronomy and Space Science, Nanjing University, 163 Xianlin Avenue, Nanjing 210023, China}
\affiliation{Key Laboratory of Modern Astronomy and Astrophysics, Nanjing University, Ministry of Education, Nanjing 210023, China}

\author[0000-0002-7440-1080]{Eda Gjergo}
\affiliation{School of Astronomy and Space Science, Nanjing University, 163 Xianlin Avenue, Nanjing 210023, China}
\affiliation{Key Laboratory of Modern Astronomy and Astrophysics, Nanjing University, Ministry of Education, Nanjing 210023, China}

\author[0000-0002-6506-1985]{Xiaoting Fu}
\affiliation{Purple Mountain Observatory,Chinese Academy of Sciences, Nanjing210023, China}



\begin{abstract}

The Milky Way is a dynamic and evolving system shaped by numerous merger events throughout its history. These mergers bring stars with kinematic and dynamic properties differing from the main stellar population. However, it remains uncertain whether any of the Galactic supernova remnants can be attributed to such a merger origin.
In this work, we compare the progenitor of Kepler's supernova to its nearby stars, ``alien'' stars, and in-situ Milky Way stellar populations. We uncover the abnormal kinematics and dynamics of Kepler's supernova and propose that its progenitor may have an extragalactic origin. We call the Type Ia supernovae (SNe Ia) produced by stars accreted into the Milky Way through merger events ``alien SNe Ia'' since they are cosmic immigrants. We estimate the rate of alien SNe Ia exploded recently using two methods: through galactic chemical evolution, and through a method without considering exact star formation history, introduced for the first time in this paper. We consider the past accretion of a few major satellite galaxies -- Kraken, Gaia-Enceladus-Sausage, the Helmi streams, Sequoia, Sagittarius, Wukong/LMS-1, and Cetus -- assuming these were dry mergers. The first method yields $1.5\times 10^{-5} - 5.0\times10^{-5}\rm\,yr^{-1}$, while the second method yields a comparable ${3.1}^{+1.8}_{-{1.1}}\times10^{-5}\rm\,yr^{-1}$ as the rate estimates for recent alien SNe Ia. These estimates represent lower bounds because we assumed no postmerger star formation.

\end{abstract}



\section{Introduction} \label{sec:intro}

Galaxies host complex and evolving ecosystems in which diverse phenomena rarely manifest in isolation. Some processes, such as formation and evolution of massive stars, are primarily governed by the present-day physical properties of their environment because of the short life time of these stars. In contrast, others--including dynamics, the evolution of long-lived stars, and the occurrence of Type Ia supernovae (SNe Ia)--are shaped by conditions and events set in motion billions of years ago,  offering insights into a galaxy's distant past. 

In the case of our Milky Way, the past decade has yielded an unprecedented depth of insights into its stellar dynamics and population composition, thanks to 
the high-precision astrometric and photometric data provided by the Gaia mission \citep{2016A&A...595A...1G} and large spectroscopic surveys on the ground such as the Large Sky Area Multi-Object Fiber Spectroscopic Telescope \citep[LAMOST;][]{2012RAA....12.1197C}, 
 H3 \citep[``Hectochelle in the Halo at High Resolution'';][]{2019ApJ...883..107C}, Gaia-ESO survey \citep{ges1, ges2}, the Apache Point Observatory Galactic Evolution Experiment \citep[APOGEE;][]{apogee}, the Galactic Archaeology with HERMES \citep[GALAH;][]{galah}, and the RAdial Velocity Experiment \citep[RAVE;][]{rave}.

Of particular relevance is the discovery of numerous substructures within the Milky Way’s stellar populations, the most notable being Gaia-Enceladus-Sausage \citep[e.g.,][]{2018Natur.563...85H, 2018MNRAS.478..611B}. Dispersed throughout the Milky Way’s inner halo and thick disk, Gaia-Enceladus-Sausage stars exhibit a distinct kinematic signature in velocity space, suggesting an origin in a disrupted galaxy that was later accreted by the Milky Way’s Main~Progenitor.
The study of substructures such as Gaia-Enceladus-Sausage is helping to disentangle which stars formed in situ and which were accreted, both within the halo \citep[see e.g.][]{Haywood_2018} and through analyses of open clusters \citep[see e.g.][]{Fu_2022}. Some of these substructures have been further explored in the context of the Milky Way’s accretion history \citep[e.g.,][]{2020ApJ...901...48N, 2022ApJ...926..107M, 2024MNRAS.527.9892Y}.

Supernovae (SNe) play an important role in regulating the chemical and dynamical evolution of galaxies. In our Milky~Way, the supernova (SN) rate is around 2 -- 3 per century \citep{1994ApJS...92..487T}, and we have known 300 -- 400 supernova remnants (SNRs) that were produced by the past SNe in the past $10^5$ -- $10^6$ yr \citep[][]{2012AdSpR..49.1313F, 2019JApA...40...36G} \footnote{\url{ http://snrcat.physics.umanitoba.ca}}$^,$ \footnote{\url{http://www.mrao.cam.ac.uk/surveys/snrs/}}. \citet{1994ApJ...437..781F} estimates the typical radio lifetime of SNRs is $60$ kyrs. However, we note that the lifetime is highly uncertain, depending on the environmental density and SNR energy. Galactic SNRs provide nearby and valuable targets to study the SN property and its feedback to galaxy. 

Here, we focus on SNe Ia
, thermonuclear explosions of white dwarfs (WDs) in binary systems, together with their remnants. 
While most SNe Ia are thought to originate from in-situ stellar populations, it is plausible that some are produced by stars accreted into the Milky Way during merger events. We refer to these events as ``alien SNe Ia'', highlighting their extragalactic origin.
We expect only alien SNe Ia and not alien core-collapse SNe because the longest delay time of core-collapse SNe is less than $\sim 0.3$ Gyr even after accounting for binary interaction \citep{2017A&A...601A..29Z} while the youngest satellite galaxy merger event in our study -- Sagittarius is accreted about 7 Gyr ago \citep{2020MNRAS.498.2472K}. 

For a single stellar population (SSP), which formed in a single episode of star formation, the delay-time distribution (DTD) describes how events -- in this case, SNe Ia -- are distributed as a function of time since the formation of an SSP. The DTD of SNe Ia has been investigated from multiple perspectives, including population synthesis models, binary dynamics, and stellar evolution, all considered in light of observational constraints \citep[e.g.,][]{Greggio2005}. When assuming a power-law DTD as a function of time, it is found that its slope is close to  $\sim t^{-1}$  \citep[e.g.,][]{2017ApJ...848...25M, 2021MNRAS.502.5882F, 2021MNRAS.506.3330W}.

In this work, we consider seven merged satellite galaxies,
i.e., Kraken \citep{2020MNRAS.498.2472K}, Gaia-Enceladus-Sausage \citep{2018MNRAS.478..611B, 2018Natur.563...85H, 2018ApJ...863L..28M} , the Helmi streams \citep{1999Natur.402...53H}, Sequoia \citep{2019MNRAS.488.1235M}, Sagittarius \citep{1994Natur.370..194I}, Wukong/LMS-1 \citep{2020ApJ...901...48N, 2020ApJ...898L..37Y}, and Cetus \citep{2009ApJ...700L..61N}. 
These accreted components have distinct kinematic signatures, enabling us to disentangle in-situ and accreted stars. If SNe Ia originate from these substructures, their remnants may also exhibit distinct kinematic and spatial properties.

Kepler’s supernova (SN 1604), the most recent historical supernova in the Milky Way, is a possible candidate for an alien SN Ia. It stays high above the Galactic plane and its progenitor star escaped the Galactic plane with a high velocity \citep[$\sim 180\, d_{4.5}\, \km\ps$,][]{1987ApJ...319..885B}\footnote{$d_{4.5}$ is defined as Kepler's distance scaled by $4.5\,\rm kpc$.}. This high velocity likely causes the highly asymmetric morphology of the SNR,
which is unusual among Type Ia SNRs \citep{2012A&A...537A.139C}. 
This raises the possibility that its progenitor originated from an accreted satellite galaxy.

 The main goal of this work is to reveal the abnormal kinematic and dynamic state of Kepler's progenitor, suggesting that Kepler is an alien SN Ia candidate, and estimate the rate and number of alien SNe Ia exploded over the past 60 kyrs. In Section \ref{sec:Data}, we describe the data used to quantify the kinematics and dynamics of Kepler's progenitor and its surrounding stars, as well as the literature data of accretion times and stellar masses of merged satellite galaxies. In Section \ref{sec:Computation of Actions and Energies}, we compute the actions and energies for Kepler's progenitor and its surrounding stars. 
 Section \ref{sec:Abnormal Kinematic and Dynamic State of Kepler's progenitor} suggests that Kepler could be an alien SN Ia candidate by comparing its progenitor's kinematics and dynamics with those of its nearby stars and known substructures of the Milky Way. In Section \ref{sec:Methods}, we introduce two complementary methods to estimate the rate and number of recent alien SNe Ia. Section \ref{sec:comparison} shows the results of estimated rate and number of recent alien SNe Ia by the two methods introduced in Section \ref{sec:Methods} and compares the differences between these different methods. The discussion and conclusion are presented in Sections \ref{sec:Discussion & Conclusion}.

\section{Data} \label{sec:Data} 

In this section, we describe the datasets used to analyze the kinematics and dynamics of Kepler’s progenitor and its surrounding stars, as well as the accretion times and stellar masses of merged satellite galaxies from the literature.

\subsection{Kinematics of Kepler's Progenitor}

Kepler is located far above the Galactic plane, with optical knots believed to trace the kinematics of its progenitor. These knots are dense and nitrogen-rich, distinguishing them from the surrounding interstellar medium and suggesting that their motion closely reflects that of the progenitor \citep{1987ApJ...319..885B, 1991ApJ...366..484B, 2012A&A...537A.139C}.

Proper motion measurements of these knots have been made in two major studies. 
\citet{1977ApJ...218..617V} found $\rm{pmRA}=-4.1\pm 1.9\, \mas \pyr, \, \rm{pmDE}=10.9 \pm 1.8 \,\mas \pyr$, with the Hooker 2.5 m telescope, the Hale 5 m telescope, and the 4 m telescope of the Cerro Tololo Inter-American Observatory, covering the period 1942 -- 1976. \citet{1991ApJ...374..186B} found $\rm{pmRA}=-6.23\pm 0.45\,\mas \pyr, \, \rm{pmDE}=4.84 \pm 0.49 \,\mas \pyr$, with the Hooker 2.5 m telescope, the Hale 5 m telescope, and the Danish 1.5 m telescope, covering the period 1942 -- 1989. The most likely reasons for their different results is the use of different optical knots. \citet{1977ApJ...218..617V} use 19 knots, \citet{1991ApJ...374..186B} use 50, with 16 knots being shared by both. 

We incorporate both measurements to account for uncertainties but emphasize the need for new observations to refine these results. These proper motions are regarded as the counterparts of the progenitor's pmRA and pmDE.

We recalculate the radial velocities of optical knots based on the measurements by \citet[][see their Table 3]{1991ApJ...366..484B}.  
We utilize the line centers of narrow components of H$\alpha$ in all the optical knots except for Knot D3 due to its excessive reduced $\chi^2$ ($\gg 2$). The radial velocity is derived as $-162\pm4\, \km\ps$, which is roughly consistent with those obtained by \citet[][$-185\, \km\ps$ for only Knots D49\&D50]{2003A&A...407..249S} and \citet[][$-275$ to $-140 \km\ps$]{1977ApJ...218..617V}.

\subsection{Kinematics of Surrounding Stars Using Gaia}

To contextualize Kepler’s progenitor within its local stellar environment,
we need to determine the 3D spatial position of Kepler, namely the coordinates and distance.

We take the coordinates $\alpha_{J2000}=17^h30^m41.25^s$ and $\delta_{J2000}=-21^\circ29^\prime32.95^{\prime \prime}$ for the geometry center of Kepler \citep{2008ApJ...689..231V}. 
The distance of Kepler has been measured in a number of studies \citep[e.g.][and the references therein]{2016ApJ...817...36S, 2020ApJ...893...98M}.
We adopt the value of 4.4 -- 7.5 kpc \citep{2020ApJ...893...98M}, which is calculated from the measured radial velocities of X-ray knots and their angular distance to the SNR center. 

In the latest Gaia Data Release 3 (DR3) catalog \citep[][]{2023A&A...674A...1G}, we search for stars within a $1.^\circ5$\footnote{A large enough sample of stars is required, with a search radius that doesn't get too large to lose meaningful comparison to Kepler's progenitor. At 5 kpc, $1.^\circ5$ corresponds to $\sim 130$ pc, which seems like a suitable option.} radius centered around the aforementioned geometric center of Kepler, using both geometric and photogeometric distance estimations \citep{2021AJ....161..147B} as distance constraints. 
We select stars by imposing the 1$\sigma$ ranges both distance estimations lie between  4.4 and 7.5 kpc.
To attenuate the impact from potential 5$\sigma$ outliers in parallax or proper motion and from potential non-single objects, we refer to the criteria in \citet{2021A&A...649A...5F}, i.e., $\text{ruwe} < 1.4$ and $\text{ipd\_frac\_multi\_peak} \le 2$ and $\text{ipd\_gof\_harmonic\_amplitude} < 0.1$ in the Gaia DR3 catalog. This resulted in a sample of  $5,348$ stars, of which  $3,507$ have radial velocity measurements from Gaia DR3. When there is radial velocity, we refer to \citet{2023A&A...674A...5K} for correction on the zero point of radial velocity.
We call these selected stars ``Kepler's surrounding stars'' or ``Kepler's nearby stars''.

\subsection{Accretion Times and Stellar Masses of Merged Satellite Galaxies}\label{sec:literature}

We adopt accretion redshifts ($z_{\rm acc}$) and stellar masses at that moment ($M_{\rm acc}$) for Kraken, Gaia-Enceladus, the Helmi streams, Sequoia and Sagittarius from \citet{2020MNRAS.498.2472K}, who derive these quantities by exploiting the ages, metallicities, and orbital properties of globular clusters formed in each merged satellite galaxies. Their $z_{\rm acc}$ and $M_{\rm acc}$ agree with other researches \citep[e.g.,][]{2012MNRAS.422..207N, 2018Natur.563...85H, 2019A&A...625A...5K, 2019MNRAS.488.1235M}.
 For Wukong/LMS-1 and Cetus, we employ the star formation history (SFH) truncation redshifts ($z_{\rm trunc}$) and stellar masses ($M_{\rm \star}$) from \citet[][see their Table 1 for details]{2022arXiv220409057N}. Because \citet{2022arXiv220409057N} does not provide the uncertainties for $\log M_{\star}$ and $z_{\rm trunc}$, we follow \citet{2020MNRAS.498.2472K} and assign a random uncertainty of 0.15 dex and a systematic uncertainty of 0.3 dex to $\log M_{\star}$, and a random uncertainty of 0.3 and a systematic uncertainty of 0.25 to $z_{\rm trunc}$. For the sake of consistent terminology, we will refer to the redshifts and stellar masses as $z_{\rm acc}$ and $M_{\rm acc}$, respectively. The adopted $z_{\rm acc}$ and $M_{\rm acc}$ are summarized in Table \ref{tab:zaccmacc}. We convert the redshifts to times ($t_{\rm acc}$) using \citet{2014A&A...571A..16P} cosmology. Throughout the text, ``Main Progenitor'' refers to the most massive progenitor galaxy of the Milky Way.

\begin{deluxetable}{lcc}
\tablecaption{Summary of the accretion redshift ($z_{\rm acc}$) and stellar mass ($\log_{10} M_{\rm acc}$) for the seven satellite galaxies\citep{2020MNRAS.498.2472K, 2022arXiv220409057N}.\label{tab:zaccmacc}}
\tablehead{
  \colhead{Satellite Galaxy} & 
  \colhead{$z_{\rm acc}$} & 
  \colhead{$\log_{10} M_{\rm acc}$} \\
  & \colhead{} & \colhead{[M$_\odot$]}
}
\startdata
Kraken          & $2.26 (0.420, 0.25)$ & $8.28 (0.175, 0.3)$ \\
Gaia-Enceladus  & $1.35 (0.245, 0.25)$ & $8.43 (0.155, 0.3)$ \\
Helmi streams   & $1.75 (0.395, 0.25)$ & $7.96 (0.185, 0.3)$ \\
Sequoia         & $1.46 (0.170, 0.25)$ & $7.90 (0.110, 0.3)$ \\
Sagittarius     & $0.76 (0.205, 0.25)$ & $8.44 (0.215, 0.3)$ \\
Wukong/LMS-1    & $0.90 (0.300, 0.25)$ & $7.10 (0.150, 0.3)$ \\
Cetus           & $2.30 (0.300, 0.25)$ & $7.00 (0.150, 0.3)$ \\
\enddata
\tablecomments{
   The values are presented in the format: $\mathrm{value\ (random\ uncertainty,\ systematic\ uncertainty)}$.
}
\end{deluxetable}

\section{Kepler's SN: An Alien Type Ia Supernova Candidate}

\subsection{Kinematics and Dynamics of Kepler's Progenitor and its Surrounding Stars}
\label{sec:Computation of Actions and Energies} 
Actions and energies, as kinematic and dynamic properties, are widely used to identify substructures in the Milky Way. In this section, we compute these quantities for Kepler's progenitor and its surrounding stars.

The accreted satellite galaxies may be disrupted and dissolved in the Milky Way after several Gyrs, but their member stars' dynamical energy and the angular momenta
remain preserved for a long time. Alien and in situ stars can be separated by inspecting these physical quantities \citep[see detail discussions in ][]{helmi2000}. We hereafter explore whether Kepler can be distinguished from stars born in the Milky~Way in $(E, j_r, j_p, j_z)$ space, where  $E$ (in $\km^2 \s^{-2}$) is the total energy including potential energy and kinematic energy, and actions $j_i \equiv \frac{1}{2\pi}\oint \dot{x}_i \, dx_i$ with $i = r, p, z$ (in $\km \ps \kpc$). $j_r$ indicates the eccentricity of an orbit, $j_p$ indicates the radius of an orbit, and $j_z$ is an indicator of the maximum vertical distance from the Galactic plane that a star can achieve. More detailed information about actions can be found in \citet{2008gady.book.....B}. 

We calculate $(E, j_r, j_p, j_z)$ through orbit integration which has been widely used to detect substructures in kinematic and dynamic space \citep[e.g.][]{2020ApJ...901...48N, 2022ApJ...926..107M, 2024MNRAS.527.9892Y}. For orbit calculation, we utilize spatial and velocity information 
with the Python package Galpy \citep{2015ApJS..216...29B}\footnote{\url{https://github.com/jobovy/galpy}}, which is a powerful tool designed for galactic-dynamics calculations. In detail, we employ two sets of potentials, i.e., MWPotential2014 \citep{2015ApJS..216...29B} and McMillan17 \citep{2017MNRAS.465...76M}. 
The result based on McMillan17 will be presented in Section \ref{sec:Abnormal Kinematic and Dynamic State of Kepler's progenitor}, and the differences between MWPotential2014 and McMillan17 will be discussed in Section \ref{sec:Discussion & Conclusion}. Because McMillan17 is axisymmetric, $j_p$ and the z-component (in the Galactic pole direction) of the angular momentum $L_z$ are identical, we will use $j_p$ and $L_z$ interchangeably. Additionally, we use the Staeckel Fudge \citep{2012MNRAS.426.1324B} algorithm implemented in Galpy to calculate actions. We assume that the Sun’s height with regard to the Galactic midplane is $z_\odot=14 \,\mathrm{pc}$ \citep{2017MNRAS.465...76M}, 
that the solar motion with respect to the local standard of rest is $v_\odot=(11.1, 12.24, 7.25) \, \km \ps$ \citep{2010MNRAS.403.1829S}, that the projection of the distance to the Galactic Centre onto the Galactic midplane is $R_\odot = 8.21 \, \mathrm{kpc}$ \citep{2017MNRAS.465...76M}, and that the circular velocity at the solar radius is $233.1 \, \km \ps$ \citep{2017MNRAS.465...76M}.

Until now we have the 3D spatial position of Kepler's progenitor\footnote{Actually, what we have is the 3D spatial position of Kepler. However, it has been only about 400 years since SN 1604, so the position of Kepler's progenitor is almost the same as Kepler.} and its nearby stars, 3D velocity of Kepler's progenitor, and 2D or 3D velocities of the nearby stars. However, conducting orbit integration requires 3D spatial position and 3D velocity, which means that only $3,507$ nearby stars are preserved while the rest $1,841$ have to be dismissed. We compare the distributions of proper motion (i.e., 2D velocities) between these two groups of stars, and find nearly the same distribution, so we expect no notable bias will be introduced using only $3,507$ stars with 3D velocity. In addition, to quantify uncertainties in the $(E, j_r, j_p(L_z), j_z)$ space, we randomly sample $1,000$ values from each relevant quantity (i.e., proper motion and radial velocity with Gaussian distribution, and distance\footnote{For stars, we use photogeometric distances from \citet{2021AJ....161..147B}.} with uniform distribution) and perform orbit integration $1,000$ times.

\subsection{Abnormal Kinematic and Dynamic State of Kepler's progenitor}
\label{sec:Abnormal Kinematic and Dynamic State of Kepler's progenitor} 
 
To understand whether Kepler could be an alien SN Ia candidate, we compare its progenitor's kinematic and dynamic property with its nearby stars, in situ Milky Way populations (i.e., the Galactic disk, the Galactic bulge, and their mixture), and accreted substructures. The kinematics and dynamics of in situ Milky Way populations and accreted substructures are from \citet{2022ApJ...926..107M}\footnote{\citet{2022ApJ...926..107M} could not detect Helmi \citep{1999Natur.402...53H} and Kraken \citep{2020MNRAS.498.2472K}, for which we refer to Appendix C therein. In short, they derive the total energy and actions of globular clusters and streams based on the association with Helmi and Kranken that are described by previous studies \citep{2019A&A...630L...4M, 2020MNRAS.498.2472K, 2021ApJ...909L..26B}.}.

\begin{figure*}
	\centering
		\centering
		\includegraphics[width=0.49\textwidth]{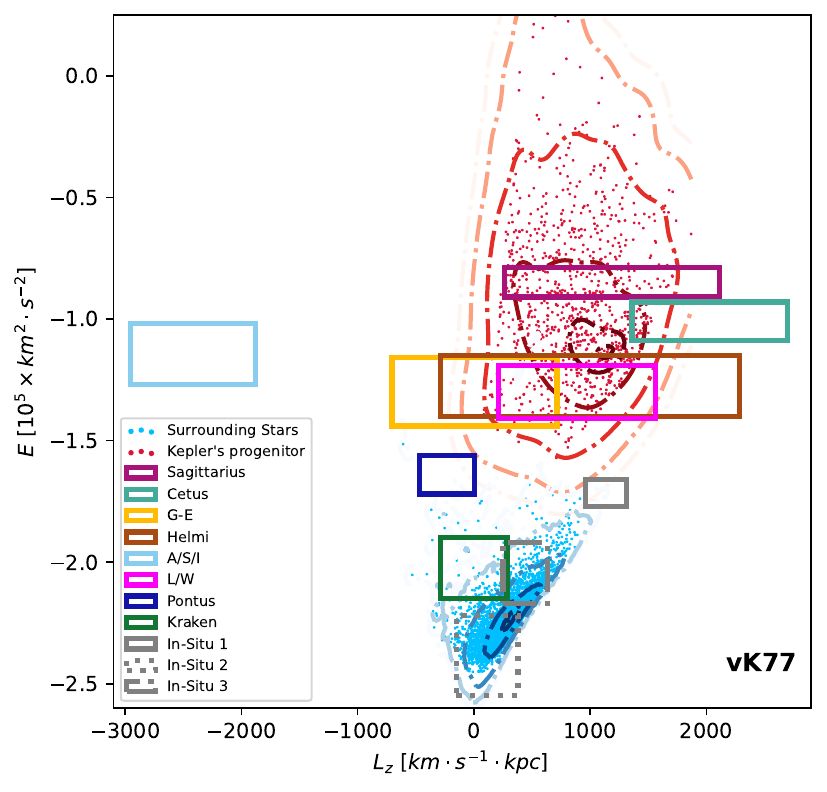}
		\includegraphics[width=0.49\textwidth]{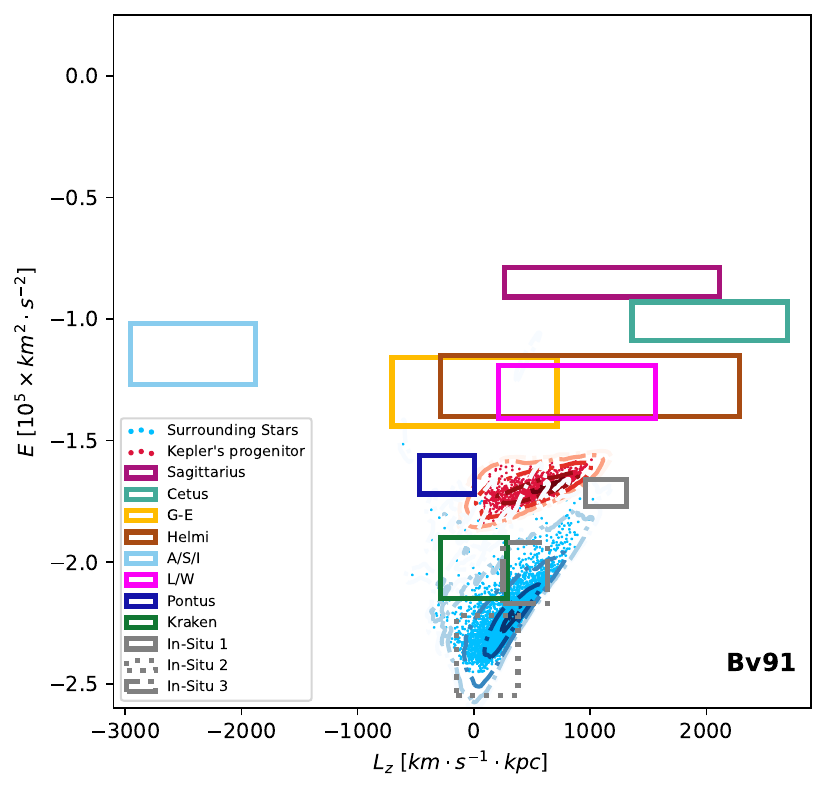}
		\caption{The distribution of Kepler's progenitor, its surrounding stars, and some known accretion events onto the Milky~Way along with in situ Milky Way populations from \citet{2022ApJ...926..107M} in the $(E,L_{z})$ space. 
        The left and right panels
        correspond to the proper motion of Kepler's progenitor, from \citet[][vK77]{1977ApJ...218..617V} and from \citet[][Bv91]{1991ApJ...374..186B}, respectively. The red color represents Kepler's progenitor, while the blue color represents its surrounding stars. The scatter points are obtained from each orbit integration (see the final paragraph in Section \ref{sec:Computation of Actions and Energies}), with only the median shown for each surrounding star. Contour lines are estimated using Gaussian kernels, with density probabilities are 0.99, 0.9, 0.5, 0.1, 0.01, and 0.001 of the maximum density probability. Dark magenta, teal, bright yellow, chestnut brown, light cyan, neon pink, navy blue, forest green, and gray boxes represent Sagittarius, Cetus, Gaia-Sausage/Enceladus (G-E), Helmi, Arjuna/Sequoia/I'itoi (A/S/I), LMS-1/Wukong (L/W), Pontus, Kraken, and in situ parts, respectively. In situ Milky Way populations contain three groups, i.e., the Galactic disk group (in situ 1), the Galactic bulge group (in situ 2), and the mixture of Galactic disk and bulge group (in situ 3) \citep{2022ApJ...926..107M}. The ranges of $E$ and $L_{z}$ for different substructures of the Milky Way are summarized in Table \ref{tab:energy_lz}.}
		\label{fig:E_Lz__scatter_contour}
\end{figure*}

\begin{deluxetable}{lcccc}
\tablecaption{Energy ($E$) and the z-component (in the Galactic pole direction) of the angular momentum ($L_z$) ranges for different Galactic substructures \citep{2022ApJ...926..107M}.\label{tab:energy_lz}}
\tablehead{
  \colhead{Substructure} & 
  \colhead{$E_{\rm lower}$} & 
  \colhead{$E_{\rm upper}$} & 
  \colhead{$L_{z,\rm lower}$} & 
  \colhead{$L_{z,\rm upper}$} \\
  & \multicolumn{2}{c}{[$10^5\ \rm{km^2\,s^{-2}}$]} & \multicolumn{2}{c}{[$\rm{km\,s^{-1}\,kpc}$]}
}
\startdata
Sagittarius & $-0.91$ & $-0.79$ & $265$ & $2115$ \\   
Cetus       & $-1.09$ & $-0.93$ & $1360$ & $2700$ \\ 
G-E         & $-1.44$ & $-1.16$ & $-705$ & $715$ \\ 
Helmi       & $-1.40$ & $-1.15$ & $-285$ & $2285$ \\ 
A/S/I       & $-1.27$ & $-1.02$ & $-2955$  & $-1880$ \\ 
L/W         & $-1.41$ & $-1.19$ & $210$ & $1560$ \\ 
Pontus      & $-1.72$ & $-1.56$ & $-470$ & $5$ \\ 
Kraken      & $-2.15$ & $-1.90$ & $-285$  & $285$ \\ 
In-Situ 1   & $-1.77$ & $-1.66$ & $960$ & $1315$ \\ 
In-Situ 2   & $-2.55$ & $-2.22$ & $-150$  & $380$ \\ 
In-Situ 3   & $-2.17$ & $-1.92$ & $250$  & $630$ \\ 
\enddata
\end{deluxetable}

\begin{figure*}
	\centering
		\centering
		\includegraphics[width=\textwidth]{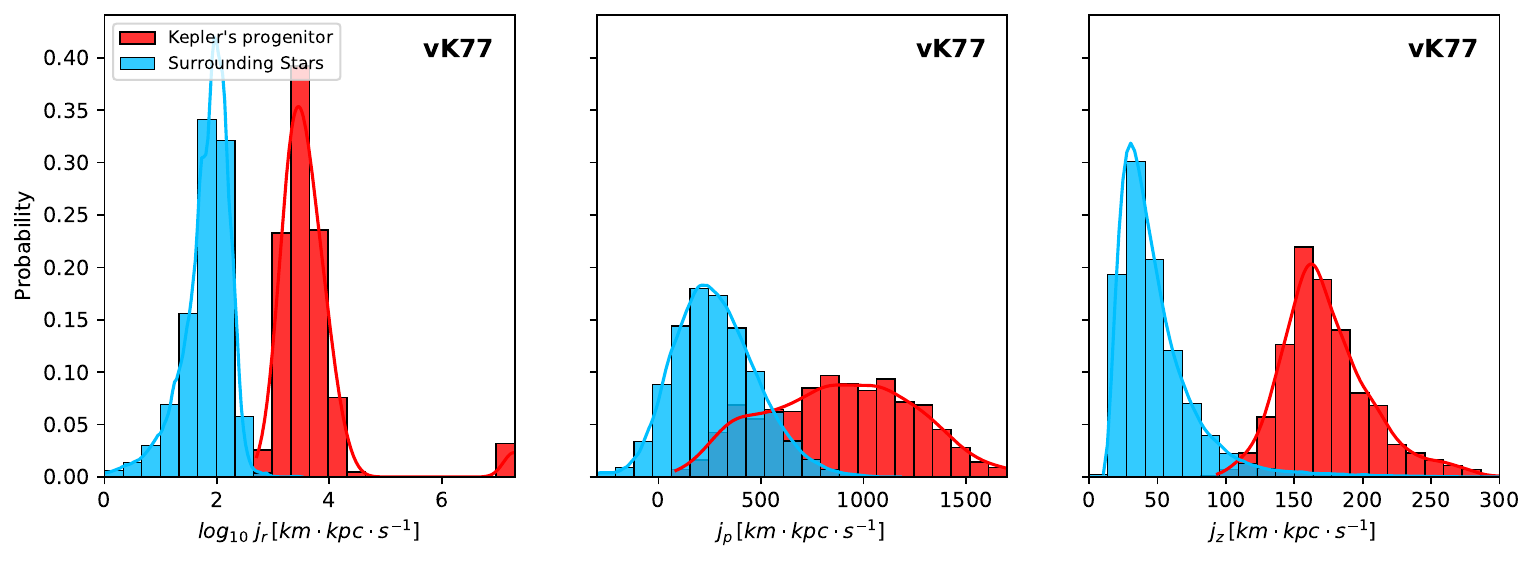}
		\includegraphics[width=\textwidth]{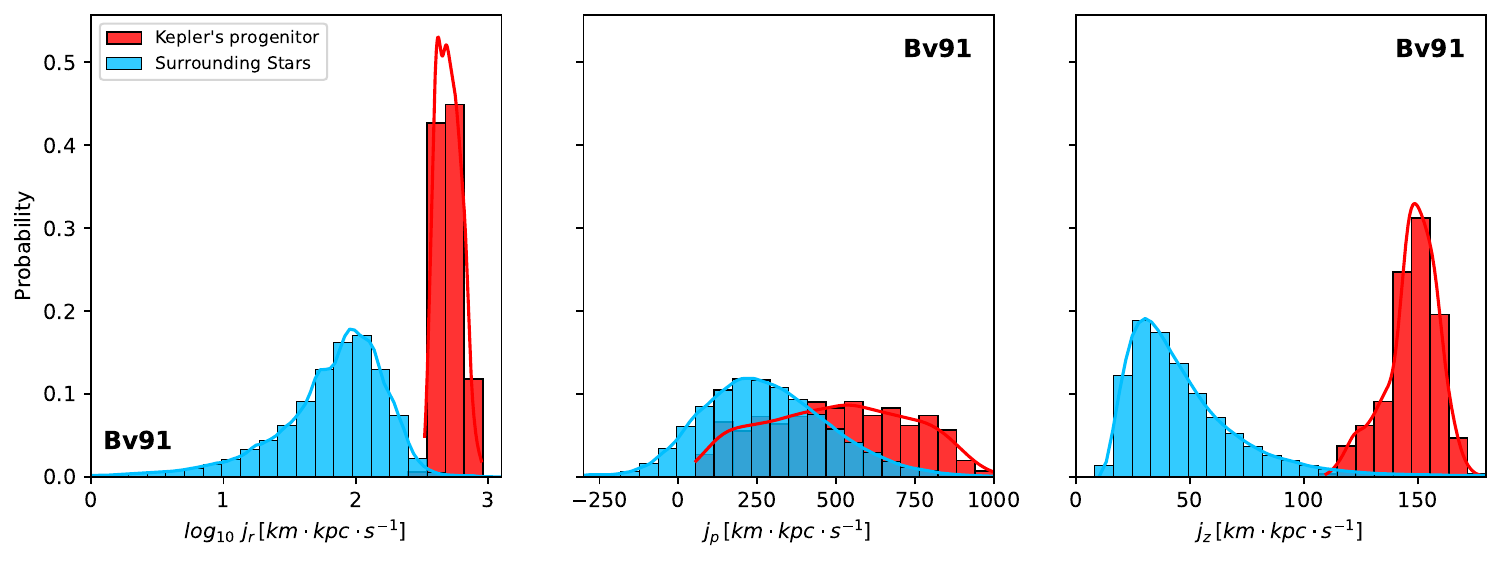}
		\caption{The 1D distribution of Kepler's progenitor and its surrounding stars in the $(j_r, j_p, j_z)$ space. Note that $j_p = L_z$. The histplots are normalized such that bar heights sum to 1. The solid lines are kernel density estimates to smooth the distributions. The top and bottom panels correspond to the proper motion of Kepler's progenitor, from \citet[][vK77]{1977ApJ...218..617V} and from \citet[][Bv91]{1991ApJ...374..186B}, respectively.}
		\label{fig:jr_jp_jz_1d}
\end{figure*}

\begin{figure*}
	\centering
		\centering
		\includegraphics[width=\textwidth]{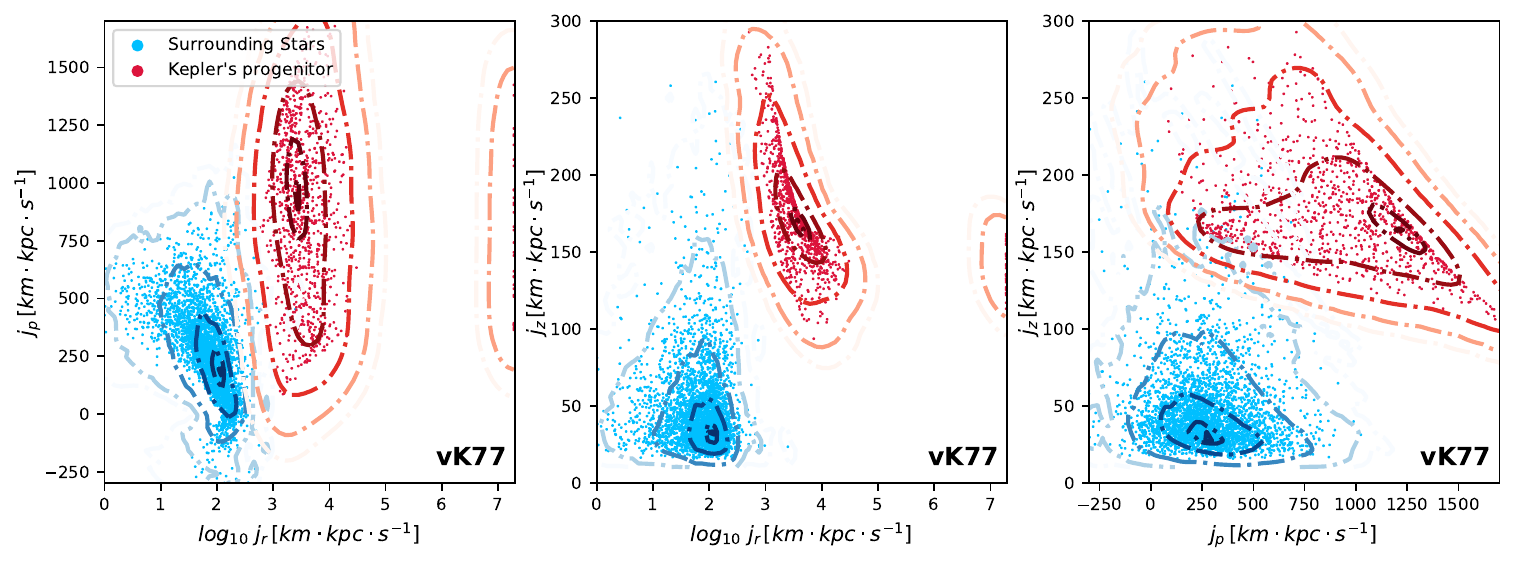}
		\includegraphics[width=\textwidth]{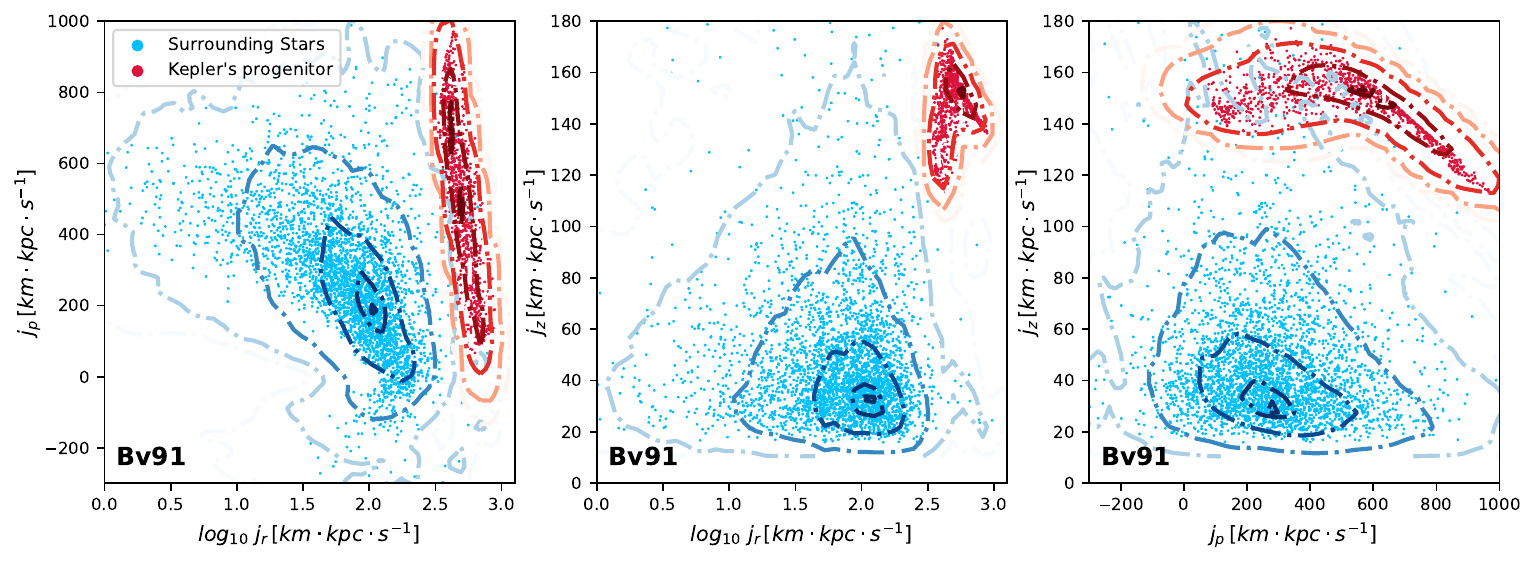}
		\caption{The 2D distribution of Kepler's progenitor and its surrounding stars in the $(j_r, j_p, j_z)$ space. Note that $j_p = L_z$. The top and bottom panels correspond to the proper motion of Kepler's progenitor, from \citet[][vK77]{1977ApJ...218..617V} and from \citet[][Bv91]{1991ApJ...374..186B}, respectively. The red color represents Kepler's progenitor, while the blue color represents its surrounding stars. The scatter points are obtained from each orbit integration, with only the median shown for each surrounding star. Contour lines are estimated using Gaussian kernels, with density probabilities are 0.99, 0.9, 0.5, 0.1, 0.01, and 0.001 of the maximum density probability.}
		\label{fig:jr_jp_jz_2d}
\end{figure*}

The distributions of Kepler's progenitor (red) and its surrounding stars (blue) in $(E, L_z)$ space (Section \ref{sec:Computation of Actions and Energies}) are depicted in Figure \ref{fig:E_Lz__scatter_contour}, where we overplot in situ Milky Way populations and some known accretion events onto the Milky~Way \citep{2022ApJ...926..107M}.
The occupied spaces of the progenitor of Kepler and its surrounding stars are completely separated, with the total energy $E$ of Kepler's progenitor significantly larger than other stars. Kepler's progenitor even has about 1\% chance of escaping the Milky~Way ($E > 0$) if we adopt the proper motion from \citet{1977ApJ...218..617V}, while virtually no opportunity to escape with the proper motion inferred from \citet{1991ApJ...374..186B}. As shown in Figure \ref{fig:E_Lz__scatter_contour}, Kepler's progenitor doesn't seem to fall into any specific region of the Galactic components identified by \citet[][in situ 1--3 stands for the Galactic disk, the Galactic bulge, and the mixture of Galactic disk and bulge]{2022ApJ...926..107M}. There are indeed a few stars (blue)
close to or even within the region occupied by Kepler's progenitor (red) in $(E, L_z)$ space ($E \sim [-1.8, -1.6]\times10^5 \times {\rm km^2\cdot s^{-2}}],\, L_z \sim [0, 1000]\times \rm{km\cdot s^{-1}\cdot kpc}$) when the proper motion from \citet{1991ApJ...374..186B} is adopted, which may suggest that Kepler is not isolated but currently has siblings in its vicinity.
We notice that the comparison to accretion events is sensitive to the proper motion measurement of Kepler's progenitor. Adopting the measurement by \citet{1977ApJ...218..617V}, Kepler's position overlaps a few accretion events, such as Sagittarius, Cetus, G-E, L/W, and Helmi, while Kepler might not relate to any of the accretion events if we use the \citet{1991ApJ...374..186B} result. 
All the anomalies in the $(E, L_z)$ space suggest that Kepler’s progenitor may not belong to the in situ Milky Way population. However, the association with accreted substructures remains inconclusive due to uncertainties in proper motion and limited observational data. Future high-precision astrometry will be critical to refining this analysis.

We also display 1D and 2D distributions in $(j_r, j_p, j_z)$ space (Figure \ref{fig:jr_jp_jz_1d} and \ref{fig:jr_jp_jz_2d}), which further demonstrates that Kepler's progenitor stands out from the stars nearby. Concretely, Kepler's progenitor has the ability to move further in both $r$ and $z$ directions, i.e., larger $j_r$ and $j_z$, and it moves in a prograde orbit with slightly faster speed than its nearby stars, i.e., positive and slighter larger $j_p$.
To recap, our analyses in $(E, L_z)$ and $(j_r, j_p, j_z)$ spaces reveal significant kinematic and dynamic anomalies of Kepler's progenitor when compared to the stars in its vicinity. In particular, the progenitor displays faster prograde motion, higher energy, and the ability to travel considerable distances in both radial ($r$) and vertical ($z$) directions with respect to Galactocentric frame.

\section{Expected rate and number of recent Alien SNe Ia}

\subsection{Methods}
\label{sec:Methods} 

To estimate the rate and number of alien SNe Ia in the Milky Way, we employed two complementary methods: (1) galactic chemical evolution models and (2) the delta function approximation (DFA).

In either way one need to derive the predicted rate and number of SNe Ia during a given time interval with SFH and the DTD of SNe Ia, where the SFH describes how star formation rate (SFR) changes over time, typically measured in solar masses per year ($\mathrm{M_\odot\,yr^{-1}}$), and the DTD of SNe Ia describes the distribution of the time intervals between the formation of a hypothetical stellar population of unit mass and the eventual SNe Ia explosion, typically measured in per solar mass per year ($\mathrm{M_\odot^{-1}\,yr^{-1}}$). We only consider the alien stars born before each satellite galaxy merged into the Milky~Way, i.e., $t_{\rm acc}$, because we poorly know the SFH after $t_{\rm acc}$. For each satellite galaxy, the contribution to the number of recent alien SNe Ia is

\begin{equation}
\begin{split}
    	N_{\rm SN-Ia}&=\int_{t_{\rm c}-t_{\rm SNR}}^{t_{\rm c}} d\tau \int_{0}^{t_{\rm acc}} dt \, {\rm SFH}(t)\, {\rm DTD}(\tau-t) \\
        &= \int_{t_{\rm c}-t_{\rm SNR}}^{t_{\rm c}} d\tau \, R_{\rm SN-Ia}(\tau) 
\end{split}
	\label{eq:SFH_convolve_DTD}
\end{equation}
where $t_{\rm c}$ is the cosmic time in an assumed cosmology, $t_{\rm SNR}$ is the typical lifetime of SNRs, $t_{\rm acc}$ marks the point where the satellite galaxy is accreted by the Milky~Way,
and $R_{\rm SN-Ia}(\tau)$ represents the rate of alien SNe Ia at cosmic time $\tau$. Note that we define the starting point of time as the moment of the Universe's birth, meaning $t_{\rm c}$ corresponds to approximately 13.8 Gyr \citep{2014A&A...571A..16P}.

\subsubsection{Galactic Chemical Evolution}
\label{sec:Method I} 

In our approach, we consider seven Milky Way stellar streams whose stellar masses and accretion redshifts onto the Milky Way are summarized in Table~\ref{tab:zaccmacc}. Their distinctive kinematic, dynamical, and chemical signatures strongly indicate these streams originated as independent satellite galaxies, later accreted by the Milky Way Main Progenitor. Consequently, we model their evolution as isolated dwarf galaxies until their merger epoch, at which point star formation is truncated due to gas stripping by the Main Progenitor. Although these mergers might have triggered some star formation, their relatively low mass represents negligible perturbations, effectively “noise”, in the Main Progenitor’s evolution, whose average behavior is accurately captured by one-zone models \citep{Matteucci_2012}.

One-zone models in fact have been successfully applied to these stellar streams, as shown for Gaia-Enceladus by \citet{Vincenzo+2019}. Additionally, one-zone models effectively reproduce many features of local dwarf galaxies \citep[e.g.,][]{Lanfranchi+2004, Lanfranchi+2007, Vincenzo+2014, Romano+2015}. Despite their simplicity, these models capture reliably the main chemical enrichment patterns observed in satellite systems. Reproducing these chemical patterns is an essential independent validity test, because abundance patterns encode the integrated SFH of a galaxy, as well as its stellar mass distribution.
Although uncertainties remain in detailed reconstructions (Yan et al. 2025, in preparation), the adopted ``dry merger'' scenario is sufficiently accurate and conservative estimate of the fraction of SNRs potentially originating from past merger events.

To perform the calculations, we employ the publicly available Galactic Chemical Evolution Model \citep[GalCEM;][]{2023ApJS..264...44G}\footnote{\url{https://github.com/egjergo/GalCEM}}, a detailed and modular code that tracks the production and evolution of isotopes within a galaxy. It implements a ``backward  algorithm": at any given time, the code uses information such as the SFR and initial mass function (IMF) to reconstruct the distribution (by mass and metallicity) of all stars expected to die within that timestep. 

The gas mass growth follows an exponential decay of primordial gas, with a timescale of formation of  7 Gyr. Further details on the model are provided in \citet{2023ApJS..264...44G}. The star~formation is then governed by a Kennicutt-Schmidt law \citep{Kennicutt_1998} with a power exponent of 1.4, typical of main-sequence galaxies. We adopt the invariant canonical IMF \citep{Kroupa_2001}. In terms of the DTD of SNe Ia, GalCEM implements the fiducial single-degenerate (SD) scenario from \citet{Greggio2005}, as explained in \citet{2023ApJS..264...44G}.  In addition, we consider the DTD from \citet{2017ApJ...848...25M}, which follows a single-power law with a slope of $-1.07$.

We first model the evolution of the Milky Way as a one-zone system \citep{Matteucci_2012}, and we treat the substructures as isolated dwarf galaxies, where star formation is halted at the redshift corresponding to their accretion onto the Main Progenitor. The eventual SNe Ia rate is governed by the galaxy’s SFR, which is, in turn, determined by the gas content at any given time. This gas content is enriched by newly ejected material from dying stars and SNe Ia, while simultaneously depleted by ongoing star formation. Once the SFR is derived from the galaxy’s evolution, the SNe Ia rate is calculated as the convolution of the SFR with the SNe Ia DTD. The DTD is normalized according to the IMF to reflect the number of stars within the mass range of 0.8 -- 8 $\rm M_{\odot}$—the progenitors of white dwarfs in binary systems—as outlined in \citet{Matteucci_1986}.

To minimize the number of free parameters, we assume hierarchical mass assembly occurs through dry mergers, meaning that during the epoch of accretion of the satellites onto the Main Progenitor, no new star formation was triggered. For all the satellites, we adopt total stellar masses and redshift of SFH truncation from the literature as explained in Sec.~\ref{sec:literature}. The chemical prescriptions for such galaxies will be analogous to that of the Main~Progenitor \citep[][ch. 7]{Matteucci_2012}, but they are rescaled to have the final stellar mass as reported in the above papers. 

\subsubsection{Delta Function Approximation}
\label{sec:Method II}

The DFA method is designed to estimate $R_{\rm SN-Ia}$ and $N_{\rm SN-Ia}$ quickly with modest precision using $t_{\rm acc}$ and $M_{\rm acc}$ instead of exact SFH under several assumptions and approximations. Our approach starts with a detailed derivation, followed by a summary of the assumptions and approximations used in the process. Ultimately, the current rate of alien SNe Ia, $R_{\rm SN-Ia}$, is derived from the ``average'' DTD times the accreted stellar mass, accounting for stellar mass loss during evolution.

We rewrite $R_{\rm SN-Ia}$ in discretization form as
\begin{equation}
	\begin{split}
		R_{\rm SN-Ia}(\tau)&= \int_{0}^{t_{\rm acc}} dt \, {\rm SFH}(t)\, {\rm DTD}(\tau-t)\\
		&= 	 \sum_{t_i \, =\, 0}^{t_{\rm acc}} \Delta t \, {\rm SFH}(t_i)\, {\rm DTD}(\tau-t_i) \\
		&= 	 \sum_{t_i \, =\, 0}^{t_{\rm acc}} M_{\rm F}(t_i)\, {\rm DTD}(\tau-t_i)
	\end{split}
	\label{eq:SFH_convolve_DTD_2}
\end{equation}
where $M_{\rm F}(t_i)$ denotes total stellar masses formed in the time interval specified by $t_i$.

Solving $R_{\rm SN-Ia}$ requires the knowledge of SFH for each satellite galaxy, which is poorly known. Therefore, we try to separate the above equation into two components, $\sum_{t_i=0}^{t_{\rm acc}}M_{\rm F}(t_i)$ and DTD$(\Delta t)$, and roughly calculate the value of $R_{\rm SN-Ia}$.
Supposing only accretion time $t_{\rm acc}$ and stellar mass of the satellite galaxy $M_{\rm acc}$ at $t_{\rm acc}$ are given\footnote{The exact SFH is generally difficult to infer. At the same time, the $M_{\rm acc}$ obtained by different studies is roughly consistent, although $M_{\rm acc}$ is closely related to SFH.}, we can draw inspiration from the closed-box model in galactic chemical evolution study, assuming no stellar mass entered or exited each satellite galaxy before $t_{\rm acc}$. This assumption is expected to hold for dwarf galaxies with stellar mass at redshift 0 smaller than $\sim 10^9 \, \rm M_{\odot}$ \citep{2017MNRAS.470.4698A}. We compare $M_{\rm acc}$ with $M_\star$ in the left panel of Figure 3 in \citet{2017MNRAS.470.4698A} at $t_{\rm acc}$ and find that $M_{\rm acc}$ are comparable to or smaller than $M_\star$ therein, so we believe that the assumption is applicable to satellite galaxies studied here. Under this assumption,  $M_{\rm acc}$ is determined by the history of star formation and stellar mass loss:
\begin{equation}
	\begin{split}
		M_{\rm acc} \, &= \sum_{t_i\,=\,0}^{t_{\rm acc}}\, M_{\rm F}(t_i)\,\left(1-{\rm Loss}\left(t_{\rm acc}-t_{i}\right)\right)
	\end{split}
	\label{eq:boundary condition}
\end{equation}
where ${\rm Loss}(t)$ represents the fraction of time-dependent stellar mass loss of a SSP through winds from asymptotic giant branch (AGB) stars and massive stars, and core-collapse SNe and SNe Ia. Since ${\rm Loss}(t)$ is a monotonically increasing function that increases slowly when $t \gtrsim 1\text{Gyr}$ \citep[with values typically spanning $0.3-0.5$; e.g.,][]{2003MNRAS.344.1000B, 2009MNRAS.399..574W, 2017PASA...34...58E}, we set a constant value ${\rm Loss}(t)=0.4$ for $t > 1 \text{Gyr}$ referring to \citet{2003MNRAS.344.1000B} assuming \citet{2003PASP..115..763C} IMF. The variation in ${\rm Loss}(t)$ arises due to differences in IMF, yields, stellar lifetime, and other aspects. Assigning ${\rm Loss}(t)=0.4$ for $t \leq 1 \text{Gyr}$ is justified in Appendix \ref{sec: appendix A}. In summary, it is reasonable to set ${\rm Loss}(t)=0.4$ for both $t > 1 \text{Gyr}$ and $t \leq 1 \text{Gyr}$.

After setting ${\rm Loss}(t)$ as a constant of 0.4, we obtain that the stellar mass in the satellite galaxy at the moment of the accretion is around 60\% of the total stellar mass formed in the past:

\begin{equation}
  \sum_{t_i\,=\,0}^{t_{\rm acc}}\, M_{\rm F}(t_i) \approx \frac{M_{\rm acc}}{0.6}
	\label{eq:mass equation}
\end{equation}

Next, we focus on the DTD component.
It is usually expressed in a power-law form \citep[e.g.,][]{2017ApJ...848...25M, 2021MNRAS.502.5882F, 2021MNRAS.506.3330W}, with units of SNe Ia rate per formed stellar mass ($\mathrm{M_\odot^{-1}\,yr^{-1}}$), 
\begin{equation}
	{\rm DTD}(\Delta t) \, =\, R \,  \left(\frac{\Delta t}{\text{Gyr}}\right)^{\alpha}
	\label{eq:DTD}
\end{equation}

where $\alpha$ typically approximate $-1$ \citep[e.g.,][]{2017ApJ...848...25M, 2021MNRAS.502.5882F, 2021MNRAS.506.3330W}, and $R$ is the DTD rate at $t=1\text{Gyr}$. From Equation \ref{eq:SFH_convolve_DTD_2} we get the minimum and maximum values of $R_{\rm SN-Ia}$: 
\begin{equation}
	\left\{
	\begin{split}
		&R_{\rm SN-Ia, min} =  \sum_{t_i \, =\, 0}^{t_{\rm acc}} M_{\rm F}(t_i)\, {\rm DTD}(\Delta t_{\rm max}) \\	 &R_{\rm SN-Ia, max} =  \sum_{t_i \, =\, 0}^{t_{\rm acc}} M_{\rm F}(t_i)\, {\rm DTD}(\Delta t_{\rm min})
	\end{split}
	\right.
	\label{eq: NIa_range_1}
\end{equation}

where $\Delta t_{\rm max} = 13.8\, \text{Gyr}$ and $\Delta t_{\rm min} =13.8\, \text{Gyr}-t_{\rm acc}$.

Then we substitute Equation \ref{eq:mass equation} into Equation \ref{eq: NIa_range_1}, arriving at
\begin{equation}
	\left\{
	\begin{split}
		&R_{\rm SN-Ia, min} =M_{\rm acc}\, {\rm DTD}(\Delta t_{\rm max})\, / \, 0.6 \\	 &R_{\rm SN-Ia, max} =M_{\rm acc}\, {\rm DTD}(\Delta t_{\rm min})\, / \, 0.6
	\end{split}
	\right.
	\label{eq: NIa_range_2}
\end{equation}

In order to find a ``middle'' value in the range of $R_{\rm SN-Ia}$, we can choose an appropriate $\Delta t$ for ${\rm DTD}(\Delta t)$ to minimize the relative distances between ${\rm DTD}(\Delta t)$ and the two extremes ${\rm DTD}(\Delta t_{\rm max})$ and ${\rm DTD}(\Delta t_{\rm min})$, ensuring that $({\rm DTD}(\Delta t) - {\rm DTD}(\Delta t_{\rm max}))/{\rm DTD}(\Delta t_{\rm max}) = ({\rm DTD}(\Delta t_{\rm min}) - {\rm DTD}(\Delta t))/{\rm DTD}(\Delta t_{\rm min})$. Setting ${\rm DTD}(\Delta t)$ equal to the harmonic mean of ${\rm DTD}(\Delta t_{\rm max})$ and ${\rm DTD}(\Delta t_{\rm min})$,  ${\rm DTD}(\Delta t_{\rm har}) = 2 \, {\rm DTD}(\Delta t_{\rm min}) \, {\rm DTD}(\Delta t_{\rm max}) / ({\rm DTD}(\Delta t_{\rm min}) + {\rm DTD}(\Delta t_{\rm max}))$, can achieve this goal. We denote this $\Delta t$ as $\Delta t_{\rm har}$.
The relative distances are 36\%, 22\%, 13\%, 17\%, 20\%, 31\%, and 12\% for Sagittarius, Gaia-Enceladus, Kraken, Helmi streams, Sequoia, Wukong/LMS-1, and Cetus, respectively, when fixing $t_{\rm acc}$ and $\alpha = -1.07$.
The ultimate formula we have is
\begin{equation}
	\begin{split}
		N^{\rm DFA}_{\rm SN-Ia}&\simeq M_{\rm acc}\, {\rm DTD}(\Delta t_{\rm har})\, \int_{t_{\rm c}-t_{\text{SNR}}}^{t_{\rm c}} d\tau \, /\, 0.6\\	 &= M_{\rm acc}\, {\rm DTD}(\Delta t_{\rm har})\, t_{\text{SNR}}\, /\, 0.6\\
		&=M_{\rm acc}\, R \,  \left(\frac{\Delta t_{\rm har}}{\text{Gyr}}\right)^{\alpha}\, t_{\text{SNR}} \, /\,0.6
	\end{split}
	\label{eq: NIa_DFA}
\end{equation}
where the superscript $\rm DFA$ is utilized for representing the method we used to derive Equation \ref{eq: NIa_DFA} because we set ${\rm Loss}(t)$ a constant (see Equation \ref{eq:mass equation}) and adopt a single value $\Delta t_{\rm har}$ for DTD component for each satellite galaxy, just like all stars being born at the same time, i.e., a delta-function-shaped SFH .

In the derivation of Equation \ref{eq: NIa_DFA}, some assumptions and approximations were used. We summarize them in Appendix \ref{sec: appendix B}. The main conclusion is that the major uncertainty is from $M_{\rm acc}$.

To implement the calculation of $N_{\rm SN-Ia}^{\rm DFA}$ using Equation \ref{eq: NIa_DFA}, we randomly sample $100,000$ times from the distribution of $\log_{10} M_{\rm acc}$, $z_{\rm acc}$, $R$, and $\alpha$ \footnote{$\alpha = -1.07^{+0.09}_{-0.09}$ and $R = 0.21^{+0.02}_{-0.02} [10^{-12}{\rm M_{\odot}^{-1}}\pyr]$ in field-galaxies \citep{2017ApJ...848...25M}.}, assuming a Gaussian Distribution except for $z_{\rm acc}$, for which we assume truncated normal distribution with a lower bound of 0. We set $t_{\text{SNR}}$ equal to $60$ kyrs \citep{1994ApJ...437..781F}, and calculate $\Delta t_{\rm har}$ (see the text under Equation \ref{eq: NIa_range_2}) by converting $z_{\rm acc}$ to $t_{\rm acc}$ using \citet{2014A&A...571A..16P} cosmology.

\subsection{Comparison of Different Alien SNe Ia Rate and Number Calculations}
\label{sec:comparison} 

In this section, we show the results calculated in Section \ref{sec:Method I} \& \ref{sec:Method II} and compare the differences between the different methods.

\begin{figure*}
     \centering
     \includegraphics[width=\textwidth]{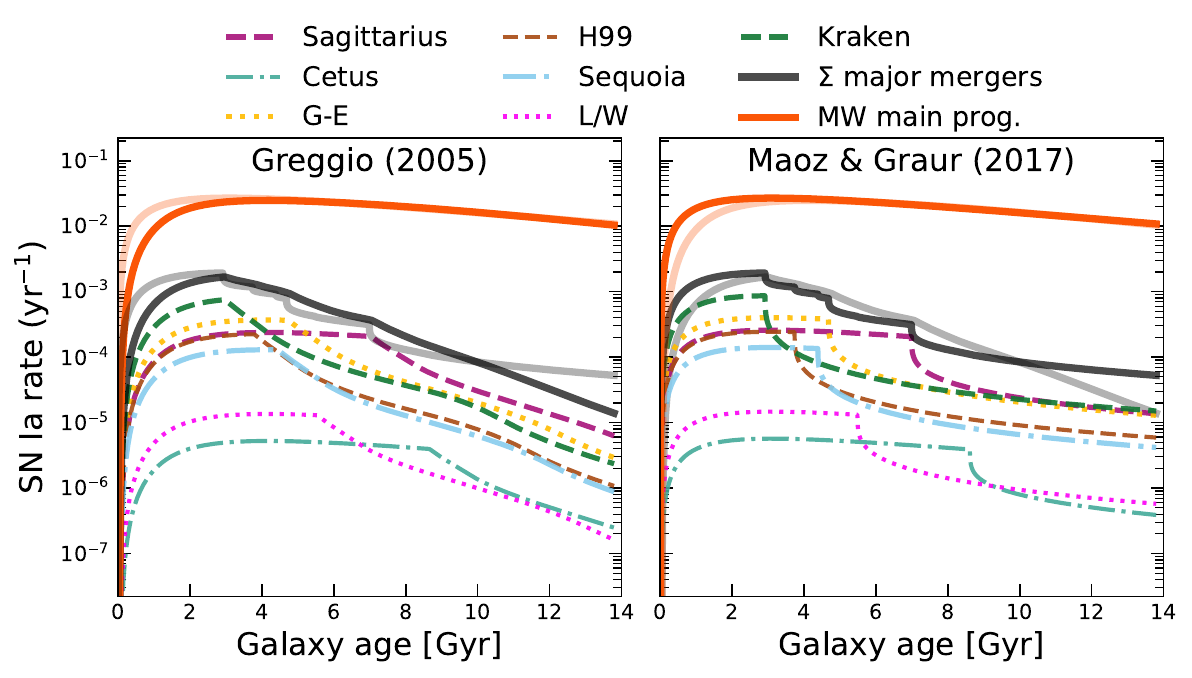}
     \caption{Evolution of the SNe Ia rates computed with GalCEM \citep{2023ApJS..264...44G} for two SNe Ia DTDs: GalCEM’s fiducial DTD \citet[][left panel]{Greggio2005} and the DTD from \citet[][right panel]{2017ApJ...848...25M}. The evolution of the Milky~Way Main~Progenitor is shown in solid orange. The remaining colored curves assume that each of the labeled substructures evolved as an isolated dwarf galaxy prior to a dry merger with the Main~Progenitor. The solid gray line represents the sum of all such mergers. For comparison, faint orange and gray transparencies in the left panel correspond to the solid lines in the right panel, and vice versa.}
     \label{fig:GalCEMrates}
 \end{figure*}

The results of Method I (see Section \ref{sec:Method I}) are shown in Fig.~\ref{fig:GalCEMrates}. We yields a rate of $1.5\times 10^{-5}\rm\,yr^{-1}$ in the case of the Greggio DTD \citep{Greggio2005}, and $5.0\times 10^{-5}\rm\,yr^{-1}$ in the case of the \cite{2017ApJ...848...25M} DTD. Integrating over the past $60$ kyrs, the typical lifetime of SNRs \citep{1994ApJ...437..781F}, we obtain 0.9 and 3.0 SN Ia events for the two DTDs, respectively. Examining the evolution of the Main Progenitor (in orange) alongside the cumulative contribution of all identified mergers considered in this study (in gray) reveals that, for both DTD descriptions, the present-day number of SNe Ia originating from the merger substructures accounts for less than 1\% of cases. When a single power-law DTD is assumed, this percentage increases by a factor of 2 to 3. This difference arises because the detailed calculations by \citet{Greggio2005} account for stellar lifetimes and evolutionary stages, producing a larger number of SNe Ia with a delay of around 1 Gyr relative to the formation of the SSP compared to the single power-law DTD from \citet{2017ApJ...848...25M}. Both the Maoz \& Graur and the Greggio DTDs are consistent with the present-day SNe Ia rate in the case of the Main~Progenitor (in orange), and they differ most at early times (i.e., in the first few Gyrs). In the case of the \citet{2017ApJ...848...25M} DTD, it is interesting to note that massive satellites merged earlier are degenerate with more recent mergers of less massive satellites. This is seen, for example, for Sagittarius, G-E, and Kraken, which all are expected to have a rate of $\sim 3\times10^{-16}\, {\rm yr}^{-1} \, \rm M_{\odot}^{-1}$ at present-day.

\begin{figure*}
	\centering
		\includegraphics[width=\textwidth]{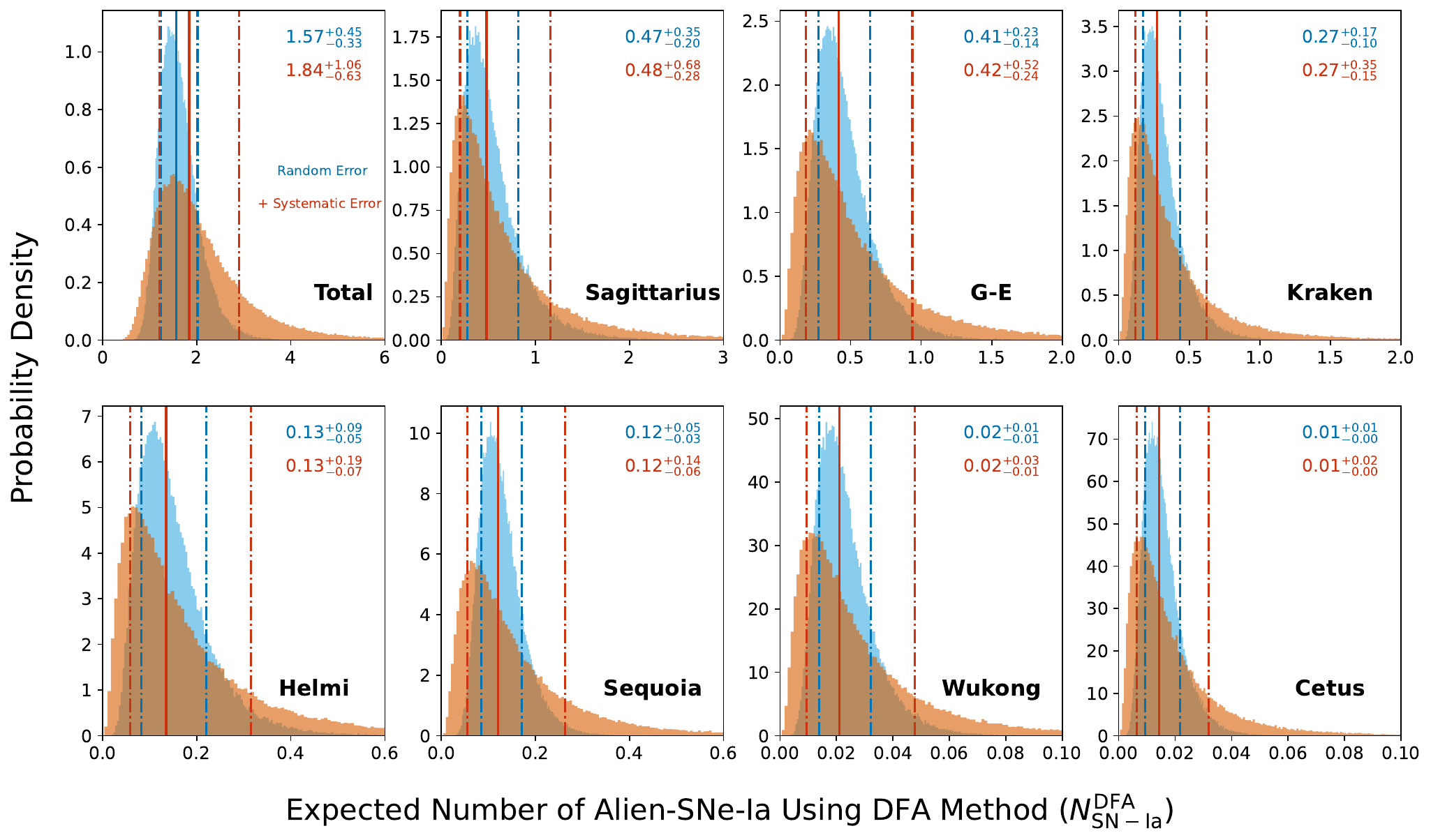}
		\caption{The $N_{\rm SN-Ia}^{\rm DFA}$ calculated from Equation \ref{eq: NIa_DFA} for each satellite galaxy and in total. The combined result for all the satellite galaxies considered are shown in the first panel, with the subsequent panels sorted by the value of $N_{\rm SN-Ia}^{\rm DFA}$ for each satellite galaxy. The blue ones only account for random uncertainties, while the orange ones also account for systematic uncertainties of 0.30 dex of $\log_{10} M_{\rm acc}$ and 0.25  of $z_{\rm acc}$ \citep{2020MNRAS.498.2472K}. The blue and orange texts in the upper-right corners are 16th-50th-84th percentiles of $N_{\rm SN-Ia}^{\rm DFA}$.}
		\label{fig:exsitu_Ia_count_each}
\end{figure*}

The $N_{\rm SN-Ia}^{\rm DFA}$ calculated through Method II (Section \ref{sec:Method II}) are shown in Figure \ref{fig:exsitu_Ia_count_each}, where the results with and without systematic error are plotted using orange and blue colors, respectively. We get $R_{\rm SN-Ia}^{\rm DFA}={3.1}^{+1.8}_{-{1.1}}\times10^{-5}\rm\,yr^{-1}$ and $N_{\rm SN-Ia}^{\rm DFA}={1.84}^{+1.06}_{-{0.63}}$ when accounting for systematic errors. Note that none of the satellite galaxies we study here have the medians of $N_{\rm SN-Ia}^{\rm DFA}$ greater than one, yet when they are combined we get $N_{\rm SN-Ia}^{\rm DFA}={1.57}^{+0.45}_{-{0.33}}$, and $N_{\rm SN-Ia}^{\rm DFA}={1.84}^{+1.06}_{-{0.63}}$ with and without systematic errors, respectively. This indicates that we expect at least one alien SNR Ia in our Milky Way if we take $60$ kyrs as the typical lifetime of SNRs.

Compared to Method I, Method II differs mainly by not incorporating the specific SFH but using the estimated minimum and maximum values (Equation \ref{eq: NIa_range_1}) to select an intermediate value for $R_{\rm SN-Ia}$. If both methods employ the Maoz \& Graur DTD, their results are essentially consistent. Since Methods I and II are relatively independent, we consider our results to be reliable.

\section{Discussion and Conclusion}
\label{sec:Discussion & Conclusion}

In this work we explore the kinematics and dynamics of Kepler’s progenitor using Gaia DR3 data around the SNR, aiming to assess its potential as an alien SN Ia and estimates the rate and number of alien SNe Ia in the Milky Way. Below, we summarize the key findings, discuss their implications, and outline directions for future work.

\begin{enumerate}
    \item 
    The kinematic and dynamic properties of Kepler’s progenitor differ significantly from the in situ Milky Way stellar populations, with higher total energy ($E$
) compared to the Galactic bulge, and lower angular momentum ($L_z$) compared to the Galactic disk. In action space ($j_r$, $j_p$, $j_z$) the progenitor exhibits greater radial and vertical motion, consistent with accreted stars from disrupted satellite galaxies.
While these anomalies suggest that Kepler’s progenitor is unlikely to belong to the in situ Milky Way population, its association with specific accreted substructures remains inconclusive. The proper motion measurements of Kepler’s progenitor introduce additional uncertainty, highlighting the need for more precise astrometric data.

    \item 
    Using two independent methods (GalCEM and DFA), we estimate the recent rate of alien SNe Ia to be  $1.5\times 10^{-5}\rm\,yr^{-1}$ to $5.0\times 10^{-5}\rm\,yr^{-1}$, corresponding to 0.9 to 3.0 events in the past 60 kyr with GalCEM, and  ${3.1}^{+1.8}_{-{1.1}}\times10^{-5}\rm\,yr^{-1}$ corresponding to ${1.84}^{+1.06}_{-{0.63}}$ events with DFA.  The results suggest that alien SNe Ia constitute a small but detectable fraction of the overall SNe Ia population, providing a unique opportunity to study the remnants of accreted satellite galaxies.

\end{enumerate}

We found that the choice of potential does not affect our qualitative results. MWPotential2014 is lighter than McMillan17, which also takes into account the gas disk. Compared to McMillan17, the stars have higher potential energy in MWPotential2014. This means that stars in MWPotential2014 can more easily escape the Milky Way. In our scenario, using the proper motion from \citet{1977ApJ...218..617V}, the escape probabilities of Kepler's progenitor for MWPotential2014 and McMillan17 are 7\% and 1\%, respectively. On the other hand, if the proper motion from \citet{1991ApJ...374..186B} is used, escape is nearly impossible for both potentials. In general, this does not affect the relative position of Kepler's progenitor to the other objects in Figures \ref{fig:E_Lz__scatter_contour}, \ref{fig:jr_jp_jz_1d}, and \ref{fig:jr_jp_jz_2d}.

In the scenario where Kepler is an alien SN Ia, the anomalous high-velocity escape of its progenitor from the Galactic plane and the asymmetric morphology of the SNR can be naturally explained.

We emphasize that the estimated rate of alien SNe Ia represents a lower bound since we have not accounted for postmerger star formation, which is poorly known and requires further investigation, and many studies have shown that a merger often triggers a temporary enhancement of star formation in the host galaxy \citep[e.g.,][]{1994ApJ...425L..13M, 2007A&A...468...61D, 2024MNRAS.527.2426A}.

There is no consensus on the progenitor system of Kepler. \citet{2012A&A...537A.139C} propose, based on hydrodynamical simulations, that the donor star in the progenitor system is an AGB star with an initial mass of $4-5 \,\rm M_\odot$, which is consistent with the kinematic and morphological properties observed in Kepler's SNR, thus favoring a SD scenario, where a WD accretes mass from a nondegenerate companion \citep{1973ApJ...186.1007W, 1982ApJ...257..780N}. Supporting this, \citet{2019ApJ...872...45S} conduct a spatially resolved X-ray spectroscopy of Kepler's SNR, and find that the estimated total hydrogen mass and Mg-to-O abundance ratio of the shocked circumstellar material can be reproduced by an AGB donor star with an initial mass matching the result of \citet{2012A&A...537A.139C}. They also report that the abundance ratios of the ejecta agree with the spherical delayed-detonation models for SNe Ia. However, \citet{2018ApJ...862..124R} find no surviving companion in Kepler's SNR, which challenges the SD scenario and instead supports either a double degenerate \citep{1984ApJS...54..335I, 1984ApJ...277..355W}\footnote{Two WDs merge.} or core-degenerate \citep{2003ApJ...594L..93L, 2011MNRAS.417.1466K} scenario \footnote{A C+O WD merges with the core of an AGB star.}. In contrast, \citet{2019MNRAS.482.5651M} suggest the surviving companion could be a subdwarf B star \citep[see also][]{2021MNRAS.507.4603M} that is below the detection limit in \citet{2018ApJ...862..124R}. In our study, we suggest that Kepler is an alien SN Ia candidate. If the progenitor formed prior to a merger event studied in this work, then Kepler's SN should have a long delay time since its progenitor formation $(> \sim 7\,\text{Gyr})$. Alternatively, if Kepler's progenitor system aligns with the \citet{2012A&A...537A.139C} scenario, then the lifetime \citep[$\sim 0.1-0.2$ Gyr;][]{1998A&A...334..505P} of the donor star with an initial mass of  $4-5 \,\rm M_\odot$ implies two possibilities: (1) its origin traces back to star formation triggered by a galaxy merger event, or (2) it formed prior to a merger event not covered in this work.

To better understand this extragalactic type of SNe and their remnants, alien SNe Ia and alien SNRs Ia, we look forward to further studies on the role of galactic merger events in shaping SFH of galaxies, and spatial, kinematic, dynamic, and chemical of the progenitors of SNRs. Given the universal prevalence of galactic merger events, it is reasonable to anticipate the presence of alien SNe Ia and alien SNRs Ia in extragalactic systems.

\begin{acknowledgments}
We thank the anonymous referee for their constructive comments. We thank J. M. Diederik Kruijssen for his thoughtful and comprehensive answers.
This work has made use of data from the European Space Agency (ESA) mission Gaia (\url{https://www.cosmos.esa.int/gaia}), processed by the Gaia Data Processing and Analysis Consortium (DPAC, \url{https://www.cosmos.esa.int/web/gaia/dpac/consortium}). Funding for the DPAC has been provided by national institutions, in particular the institutions participating in the Gaia Multilateral Agreement. W.L.H. and P.Z. acknowledge the support from National Natural Science Foundation of China (NSFC) grant No. 12273010 and the China Manned Space Program with grant no. CMS-CSST-2025-A14.
E.G. acknowledges the support of NSFC under grants NOs. 12173016, 12041305, and the Program for Innovative Talents, Entrepreneur in Jiangsu.
E.G. and X.F. acknowledges the science research grants from the China Manned Space Project with NOs. CMS-CSST-2021-A08, CMS-CSST-2021-A07.
X.F. acknowledges the support of NSFC No. 12203100. 
\end{acknowledgments}

%

\facilities{Gaia}


\software{Galpy \citep{2015ApJS..216...29B},  
          GalCEM \citep{2023ApJS..264...44G},
          NumPy \citep{2020Natur.585..357H},  
          SciPy \citep{2020NatMe..17..261V},
          Matplotlib \citep{2007CSE.....9...90H}
          }



\appendix

\section{stellar mass loss}
\label{sec: appendix A} 
For $t \leq 1 \text{Gyr}$, we calculate the averages of ${\rm Loss}(t)$ that are about 36\%/30\%/24\%/13\% over the intervals 0 to 1 Gyr/300 Myr/100 Myr/10 Myr using Equation 16 in \citet{2013MNRAS.428.3121M}, who fit the mass loss of a stellar population developed by \citet{2003MNRAS.344.1000B} with an Milky~Way canonical IMF \citep{Kroupa_2001,2003PASP..115..763C}. This means that for stars formed in the last 1 Gyr before $t_{\rm acc}$, when we consider they are all formed in the last 1 Gyr/300 Myr/100 Myr/10 Myr with uniform SFH, the estimated values of $\sum_{t_i\,=\, t_{\rm acc}-1 \rm Gyr}^{t_{\rm acc}}\, M_{\rm F}(t_i)$ are about $7\%/15\%/21\%/31\%$ lower compared to setting ${\rm Loss}(t)=0.4$. Taking 80\% (50\%, 20\%) stellar mass formed before $t_{\rm acc} - 1\rm Gyr$ and 20\% (50\%, 80\%) of that formed in the last 1 Gyr before $t_{\rm acc}$, the lower values of $\sum_{t_i\,=\, t_{\rm acc}-1 \rm Gyr}^{t_{\rm acc}}\, M_{\rm F}(t_i)$ will cause $\sum_{t_i\,=\, 0}^{t_{\rm acc}}\, M_{\rm F}(t_i)$ to be $\sim 1\%/3\%/4\%/6\%$ ($\sim 3\%/7\%/10\%/15\%$, $\sim 5\%/12\%/17\%/25\%$) lower. Hence, assigning ${\rm Loss}(t)=0.4$ for $t \leq 1 \text{Gyr}$ is justified, even in extreme scenarios.

\section{summary of assumptions and approximations in the DFA method}
\label{sec: appendix B} 
First, to circumvent the lack of precise SFH, we use $M_{\rm acc}$ instead, which is more consistent over different studies \citep[e.g.,][and references therein]{2020MNRAS.498.2472K}. Specifically, we assume no stellar mass entered or exited each satellite galaxy before $t_{\rm acc}$, whose applicability has already been discussed above, and adopt a constant value of ${\rm Loss}(t)$ in Equation \ref{eq:boundary condition}. 
Additionally, we choose $\Delta t_{\rm har}$ for $\Delta t$ in ${\rm DTD}(\Delta t)$ to minimize the maximum relative distances to both extremes, i.e., ${\rm DTD}(\Delta t_{\rm max})$ and ${\rm DTD}(\Delta t_{\rm min})$, where ${\rm DTD}(\Delta t_{\rm har})$ equal to the harmonic mean of ${\rm DTD}(\Delta t_{\rm max})$ and ${\rm DTD}(\Delta t_{\rm min})$. 

In order to estimate the uncertainties stemmed from adopting a constant value of ${\rm Loss}(t)$ and choosing $\Delta t_{\rm har}$ for $\Delta t$ in ${\rm DTD}(\Delta t)$, we look into the dynamic ranges\footnote{Dynamic range is the ratio between the largest and smallest values that a certain quantity can assume.} of $1 / (1-{\rm Loss}(t))$ (see Equations \ref{eq:boundary condition} and \ref{eq:mass equation}) and ${\rm DTD}(\Delta t)$ (see Equation \ref{eq: NIa_range_2}), which are about 1.4 and 2, respectively, while the 1 $\sigma$-dynamic-range of $M_{\rm acc}$ is about 5 when considering systematic uncertainty \citep{2020MNRAS.498.2472K}. That is to say, even if we multiply the maximum values of dynamic ranges of $1 / (1-{\rm Loss}(t))$ and ${\rm DTD}(\Delta t)$, it would only be about half of the 1 $\sigma$-dynamic-range of $M_{\rm acc}$. Hence we believe that the uncertainties from setting a constant value of ${\rm Loss}(t)$ and setting $\Delta t_{\rm har}$ for $\Delta t$ in ${\rm DTD}(\Delta t)$ are modest compared to $M_{\rm acc}$. Finally, we remind readers that Equation \ref{eq: NIa_DFA} can have wider applications than mentioned here. For instance, one can apply it to an isolated extragalactic galaxy rather than a satellite galaxy of the Milky Way, a given interval of time rather than from $t_{\text{SNR}}$ ago to $t_{\rm c}$, and needless to say, a different environment for DTD. Meanwhile, one should keep in mind that if $\Delta t_{\rm max} / \Delta t_{\rm min}$ is too large, for example, larger than 5, that is, larger than 1 $\sigma$-dynamic-range of $M_{\rm acc}$, Equation \ref{eq: NIa_DFA} will be unreliable. The reason for it can be seen in Equation \ref{eq:DTD}. Since $\alpha$ is about -1, $\Delta t_{\rm max} / \Delta t_{\rm min} \simeq {\rm DTD}(\Delta t_{\rm min})/{\rm DTD}(\Delta t_{\rm max})$, i.e., the dynamic range of ${\rm DTD}(\Delta t)$.
\bibliography{sample631}{}
\bibliographystyle{aasjournal}



\end{document}